\documentclass[two column, amssymb, amsmath,nobibnotes,nofootinbib, aps, showpacs, prb]{revtex4}
\usepackage[dvips]{graphicx}

\usepackage{graphicx}

\begin{document}
\title{$\mu$SR and inelastic neutron scattering investigations of the  noncentrosymmetric antiferromagnet CeNiC$_2$}
\author{A.Bhattacharyya$^{1,2}$}
\email{amitava.bhattacharyya@stfc.ac.uk} 
\author{D.T. Adroja$^{1,2}$} 
\email{devashibhai.adroja@stfc.ac.uk}
\author{A.M. Strydom$^2$} 
\author{A. D . Hillier$^{1}$}
\author{ J.W. Taylor$^{1}$}
\author{A. Thamizhavel$^3$} 
\author{S. K. Dhar$^3$} 
\author{W. A. Kockelmann$^{1}$}
\author{B. D. Rainford$^{4}$}
\affiliation{$^1$ISIS Facility, Rutherford Appleton Laboratory, Chilton, Didcot Oxon, OX11 0QX, UK} 
\affiliation{$^2$Physics Department, University of Johannesburg, PO Box 524, Auckland Park 2006, South Africa}
\affiliation{$^3$ Tata Institute of Fundamental Research, Homi Bhabha Road, Colaba, Mumbai 400005, India}
\affiliation{$^4$Department of Physics, Southampton University, Southampton S09 5NH , United Kingdom}
\date{\today}
\begin{abstract}
The magnetic state of the  noncentrosymmetric antiferromagnet  CeNiC$_2$ has been studied by magnetic susceptibility, heat capacity, muon spin relaxation ($\mu$SR) and  inelastic neutron scattering (INS) measurements. CeNiC$_2$ exhibits three magnetic phase transitions at $T_{N_1}$ = 20 K, $T_{N_2}$ = 10 K and $T_{N_3}$ = 2.5 K. The presence of long range magnetic order below 20 K is confirmed by the observation of oscillations in the $\mu$SR spectra between 10 and 20 K and a sharp increase in the muon depolarization rate.  INS studies reveal two well-defined crystal electric field (CEF) excitations around 8 and 30 meV. INS data have been analyzed using a CEF model and the wave functions were evaluated. We also calculated the direction and magnitude of the ground state moment using CEF wave functions and compare the results with that proposed from the neutron diffraction. Our CEF model correctly predicts that the moments order along the $b-$axis (or $y$-axis) and the observed magnetic moment is 0.687(5) $\mu_B$, which is higher than the moment  observed from the neutron diffraction (0.25 $\mu_B$/Ce). We attribute the observed reduced moment due to the Kondo screening effect.
  
\end{abstract}

\maketitle

\section{Introduction}
A rich variety of novel phenomena perceptible in Ce-based strongly correlated electron systems due to the duality between the localized and the itinerant nature of $f$-electrons, such as heavy electron and mixed valence behavior, Kondo insulator or semiconductors, unconventional superconductivity, spin and charge density waves, spin and charge gap formation and metal-insulator transition of special recent interest, is the fascinating phenomena associated with magnetic quantum criticality and non-Fermi liquid behavior arising from cooperative behavior at a zero temperature phase transition.~\cite{ag,cmv,pc,psr,grs,hvl,ams}

\begin{figure}[t]
\vskip 0.4 cm
\centering
\includegraphics[width = 9 cm]{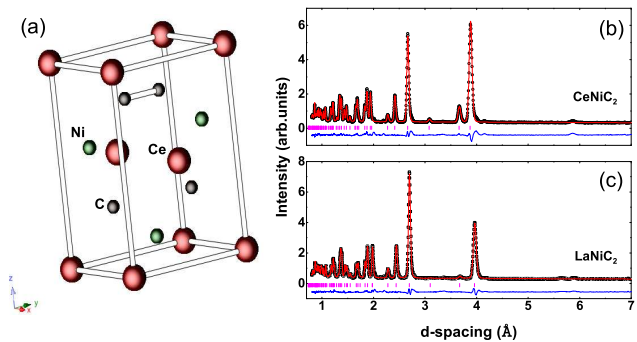}
\caption {(Color online) (a) The orthorhombic crystal structure of CeNiC$_2$ where the Ce atoms are in red, the Ni atoms are in green, and the C atoms are in black. (b) and (c) show the Rietveld refinement of the neutron powder diffraction pattern of CeNiC$_2$ and LaNiC$_2$ at 300 K. The data are shown as black circles, and the result of the refinement as solid lines (red).}
\end{figure}

\par
RNiC$_2$ (R = rare earth) compounds crystallize in the noncentrosymmetric (NCS) orthorhombic CeNiC$_2$ type structure (space group Amm2) in which a mirror plane is missing along the $c$-axis.~\cite{Onodera1998} Diverse ground states such as the superconducting, antiferromagnetic (AFM)/ferromagnetic (FM) and the charge density-wave (CDW) state are reported so far in RNiC$_2$(R = La, Ce, Pr, Sm, Gd, Tb and Er), reflecting the competition among different electronic states.~\cite{Onodera1998, K Motoya, W. Sch, M. Murase}  No structural phase transition has been reported in these compounds, and hence the spin and charge ordering phenomena in RNiC$_2$ compounds are thus native to the NCS crystal structure. LaNiC$_2$ is a superconductor below about 2.7 K.~\cite{whl, A. D. Hillier} The absence of an inversion center is believed to give rise to mixed wave pair states (s-wave singlet and p-wave triplet) in NCS superconductors. It has been argued that heavy fermion effects are not significant except perhaps for R = Ce (perhaps mixed valence effects for Sm). Murase et al ~\cite{M. Murase} found that rare-earth intermetallic compounds RNiC$_2$ with R = Nd, Sm, Gd and Tb show anomalous temperature dependences of electrical resistivity and lattice constants.~\cite{M. Murase} They proposed that these anomalies are attributed to CDW transitions. X-ray scattering studies of SmNiC$_2$ reveal satellite peaks corresponding to an incommensurate wave vector (0.5, 0.52, 0) below 148 K at which the resistivity shows an anomaly, signing development of a charge-density wave.~\cite{Shimomura} The rare earth local moments order antiferromagnetically in most of the RNiC$_2$ compounds (apart from R = Pr), and SmNiC$_2$ undergoes a first-order ferromagnetic transition at $T_C$ = 17.5 K.~\cite{Shimomura} Those results indicate that RNiC$_2$ compounds having various magnetic orders are promising candidates for the systematic investigation of the interplay of CDW and magnetic order. In this system, nickel atoms are nonmagnetic and rare-earth elements mainly contribute to the magnetic properties showing the character of local magnetic moments interacting through the Ruderman$-$Kittel$-$Kasuya$-$Yosida (RKKY) interaction. The magnetic properties of the RNiC$_2$ series are affected strongly by the orthorhombic crystalline electric field (CEF).

\par
In this paper we report a detailed investigation on CeNiC$_2$ by dc magnetic susceptibility ($\chi$), isothermal magnetization ($M$), heat capacity ($C_P$), neutron diffraction, muon spin relaxation ($\mu$SR) and inelastic neutron scattering (INS) measurements. Below 20 K a sharp anomaly is observed in our $\chi(T)$ and $C_P$ data which corresponds to a transition from the paramagnetic (PM) to AFM state. $C_P (T)$ data also show a second anomaly around 2.5 K and a further weak anomaly near 10 K. The presence of long-range magnetic order is also revealed by $\mu$SR measurements where oscillations are observed in the spectra below 20 K. A doublet ground state is inferred both from $C_P (T)$ and INS data. INS reveals two well-defined CEF excitations at 9 and 28 meV, indicating the localized nature of the $4f-$electrons in CeNiC$_2$.

\begin{table}
\begin{center}
\caption{A summary of the results obtained from the refinement of the room-temperature neutron powder diffraction data: the lattice parameters and atomic positions. The site occupancy for all the atoms was fixed to 100 \%. The reliability factors (weighted profile factor $R_{wp}$ = 5 $\%$) were calculated by comparing the fit to the data (space group no. 38, Amm2).}
\begin{tabular}{lccccccccccccccc}
\hline
&&  &&  a (\AA) && b (\AA)  && c (\AA) &&\\ 
\hline
CeNiC$_2 $&&  && 3.876(2) && 4.548(2)  && 6.161(1) && \\ 
LaNiC$_2$ && &&  3.957(2) && 4.561(1)  && 6.199(2) && \\  
\hline
 Atom &&site &&  x && y  && z  &&\\ 
\hline
Ce && 2a  && 0.00 && 0.00  && 1.00  &&\\ 
Ni && 2b  &&0.500 && 0.00 && 0.6138(2) &&   \\ 
C && 4e  &&0.500 && 0.3509(2)&& 0.8043(1)  && \\ 
\hline
\hline
La &&  2a &&0.00 && 0.00  && 1.00  &&\\ 
Ni && 2b  &&0.500 && 0.00 && 0.6137(1)  &&\\ 
C && 4e  && 0.500 && 0.3513(1) && 0.8047(2)  &&\\ 
\hline 
\hline
\end{tabular}
\end{center}
\end{table}

\begin{figure}[h!]
\vskip 0.4 cm
\centering
\includegraphics[width = 8 cm]{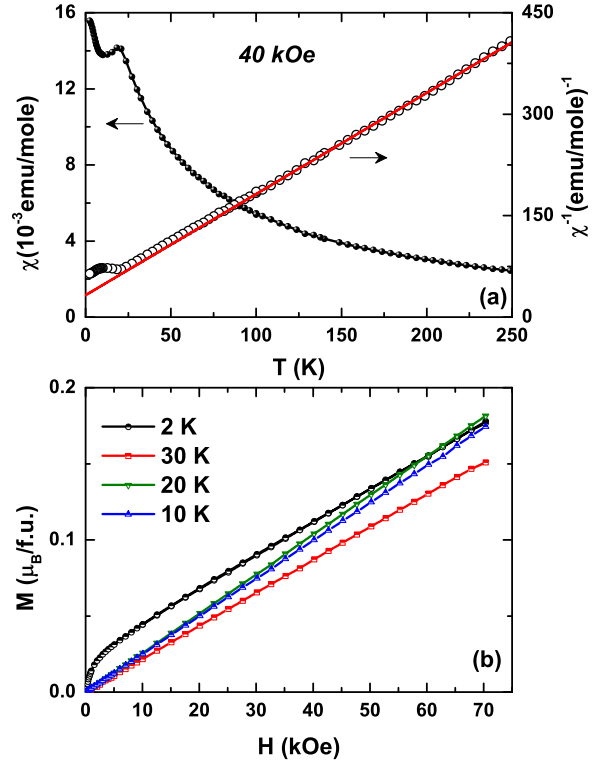}
\caption {(Color online) (a) Temperature dependence of the magnetic susceptibility and the inverse $\chi(T)$ of polycrystalline CeNiC$_2$. The straight lines show fits to CW behavior in the high-temperature region. (b) Isothermal field dependence of magnetization  at different constant temperatures.}
\end{figure}

\begin{figure}[t]
\vskip 0.4 cm
\centering
\includegraphics[width = 8 cm]{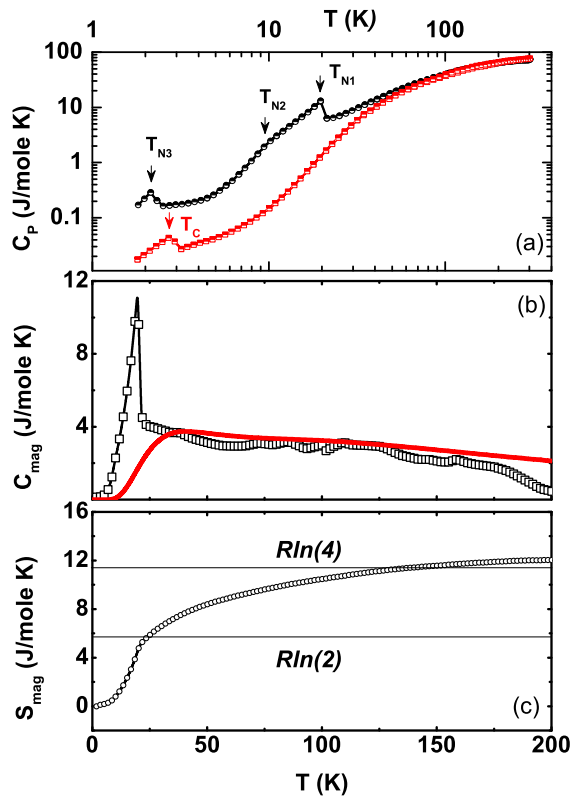}
\caption {(Color online) (a) Temperature dependence of the heat capacity of CeNiC$_2$ (black circle) and the phonon reference compound LaNiC$_2$ (red square) (b) Estimated magnetic contribution of the heat capacity plotted as $C_{mag}$ vs.\ $T$ and the red curve represents the crystal electric field contribution to specific heat according to the crystal field level scheme deduced from the inelastic neutron scattering data (c) The temperature-dependent magnetic entropy estimated from the experimental data in  (b).}
\end{figure}

\begin{figure}[t]
\vskip 0.4 cm
\centering
\includegraphics[width = 9.5 cm]{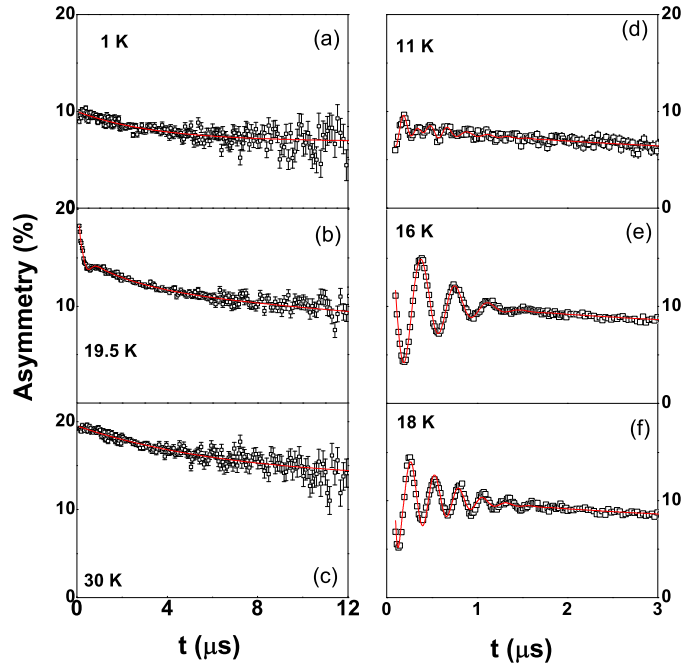}
\caption {(Color online) Zero-field $\mu$SR spectra plotted as asymmetry versus time at various temperatures of CeNiC$_2$. The solid lines depict fits using Eq. (1) (see text).}
\end{figure}

\begin{figure}[t]
\vskip 0.4 cm
\centering
\includegraphics[width = 9.5 cm]{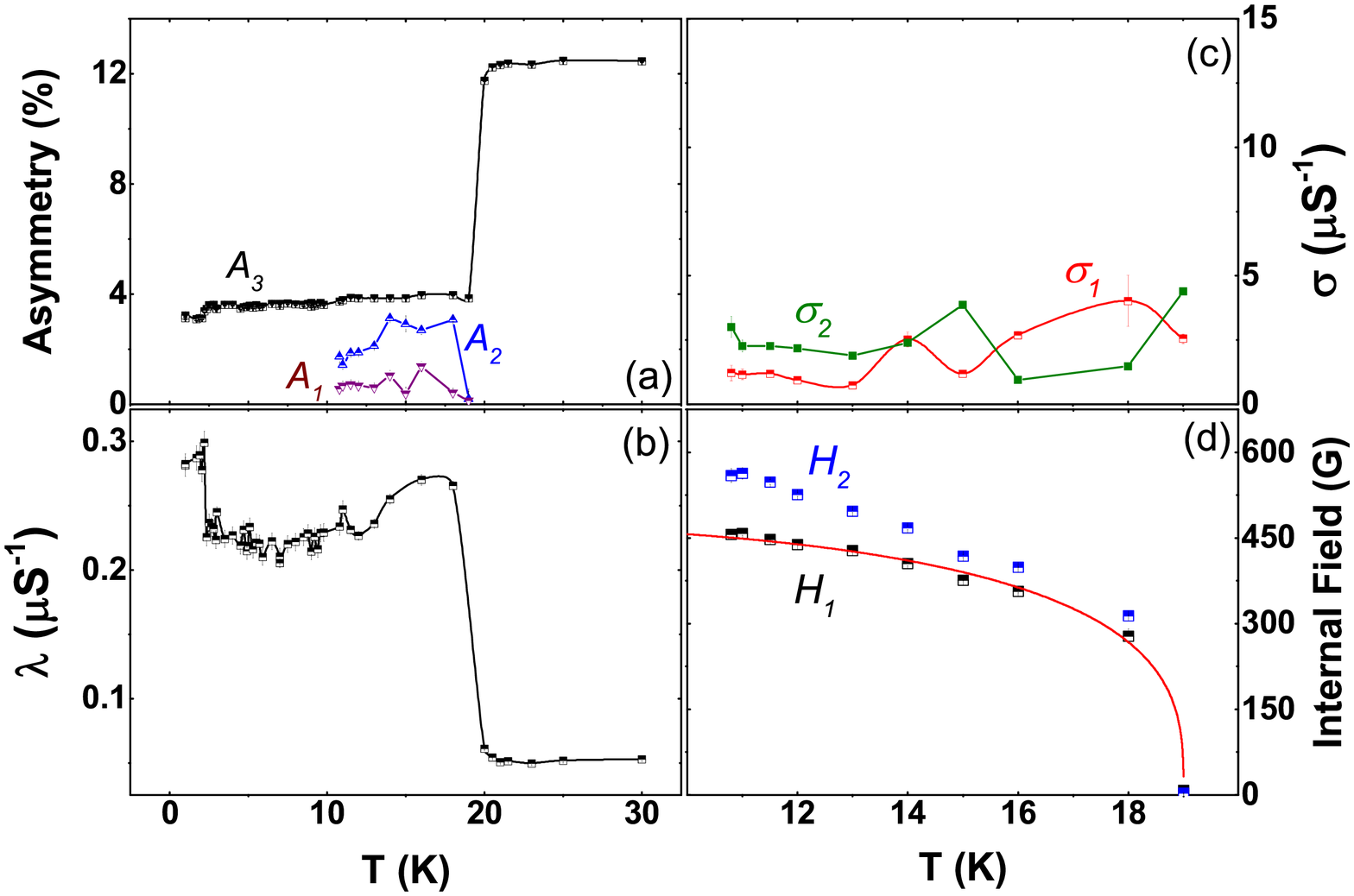}
\caption {(Color online) The temperature dependence of (a) the initial asymmetries $A_1$, $A_2$ and $A_3$  (b) the depolarization rate $\lambda$, (c) the depolarization rates $\sigma_1$ and $\sigma_2$, and (d) the internal fields $H_1$ and $H_2$. The solid line in (d) is fit to Eq. 2 (see text).}
\end{figure}

\begin{figure}[t]
\vskip 0.4 cm
\centering
\includegraphics[width = 8 cm]{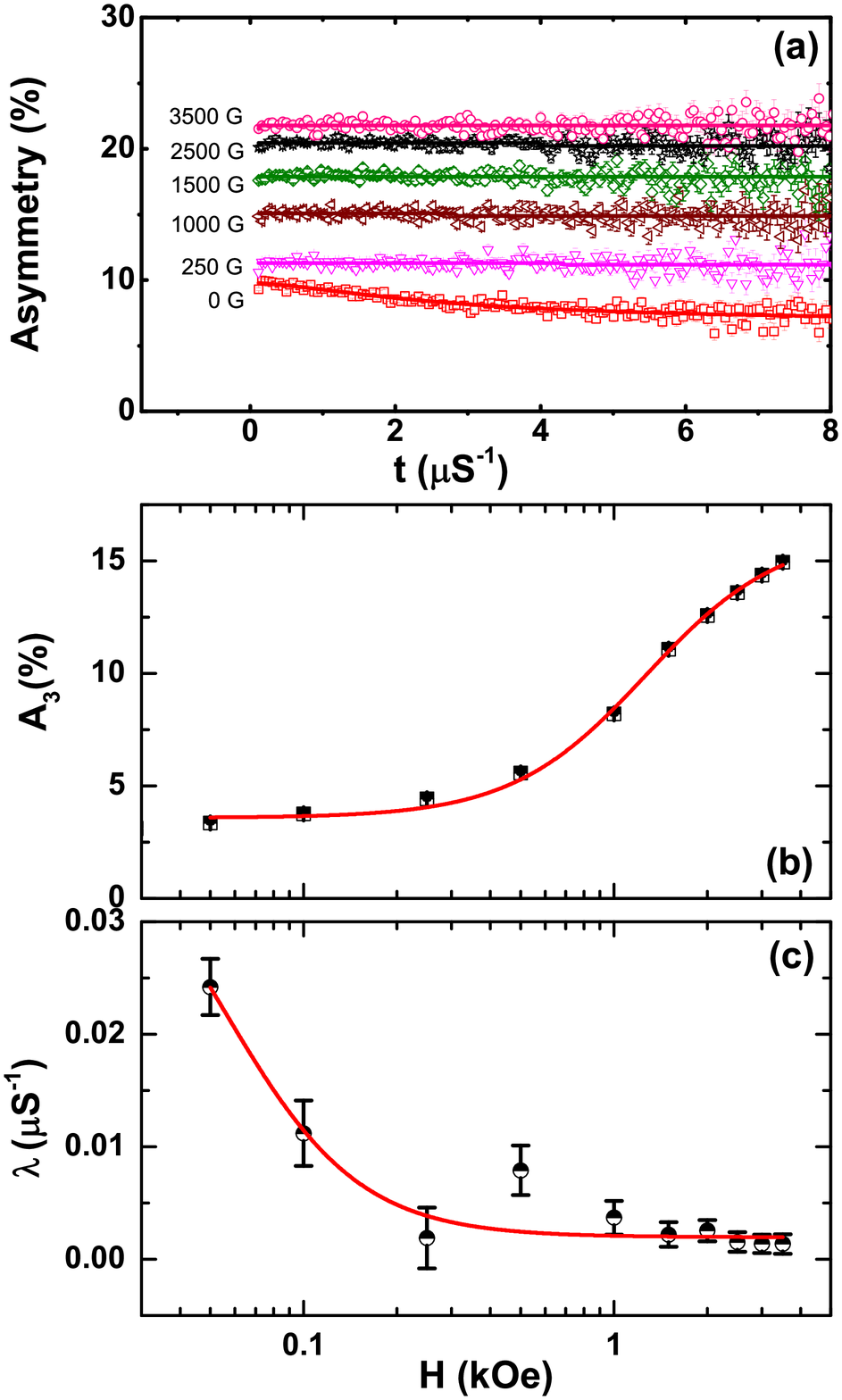}
\caption {(Color online) (a) Field dependent $\mu$SR spectra at 1.2 K . (b) the longitudinal component of the initial asymmetry ($A_3$) and depolarization rate ($\lambda$) as a function of applied magnetic field at 1.2 K. The solid lines represent fits to Eq. (3) and Eq. (4) respectively.}
\end{figure}

\section{Experimental Details}
Polycrystalline samples of CeNiC$_2$ and LaNiC$_2$ were prepared by arc-melting of the constituent elements (Ce : 99.999 wt.-\%, La : 99.999 wt.-\%, Ni : 99.999 wt.-\%, C : 99.999 wt.-\%) in an argon atmosphere on a water cooled copper hearth. After being  flipped and remelted several times, the buttons were wrapped in tantalum foil and annealed at 1000 $^0$ C for 168 h under a dynamic vacuum, better than 10$^{-6}$ Torr. Powder X-ray diffraction measurements were carried out using a Panalytical X-Pert Pro diffractometer. Magnetic susceptibility measurements were made using a MPMS SQUID magnetometer (Quantum Design). Heat capacity measurements were performed by the relaxation method in a Quantum Design physical properties measurement system (PPMS).
\par
The muon spin relaxation and inelastic neutron scattering experiments were carried out at the ISIS Pulsed Neutron and Muon Facility of the Rutherford Appleton Laboratory, United Kingdom. To check the phase purity of the samples room temperature neutron diffraction measurements were carried out using General Materials (GEM) time of flight (TOF) diffractometer. The INS measurements were carried out on the time-of-flight MARI spectrometer between 5 and 100 K. The powder samples were wrapped in thin Al foil and mounted inside a thin-walled cylindrical Al can, which was cooled down to 4.5 K inside a top-loading closed cycle refrigerator with helium exchange gas around the sample. Incident neutron energies ($E_i$) of 8, 20 and 50 meV were used on MARI selected via a Gd-Fermi chopper.  The $\mu$SR measurement was carried out on the MUSR spectrometer with the detectors in a longitudinal configuration. The powdered sample was mounted on a high purity silver plate using diluted GE varnish and covered with kapton film which was cooled down to 1.2 K in a standard $^4$He cryostat with He-exchange gas.  Spin-polarized muon pulses were implanted into the sample and positrons from the resulting decay were collected in positions either forward or backwards of the initial muon spin direction. The asymmetry is calculated by, $G_z(t) =[ {N_F(t) -\alpha N_B(t)}]/[{N_F(t)+\alpha N_B(t)}]$, where $N_B(t)$ and $N_F(t)$ are the number of counts at the detectors in the forward and backward positions and $\alpha$ is a constant determined from calibration measurements made in the paramagnetic state with a small (20 G) applied transverse magnetic field. 

\begin{figure}[b]
\vskip 0.4 cm
\centering
\includegraphics[width = 9 cm]{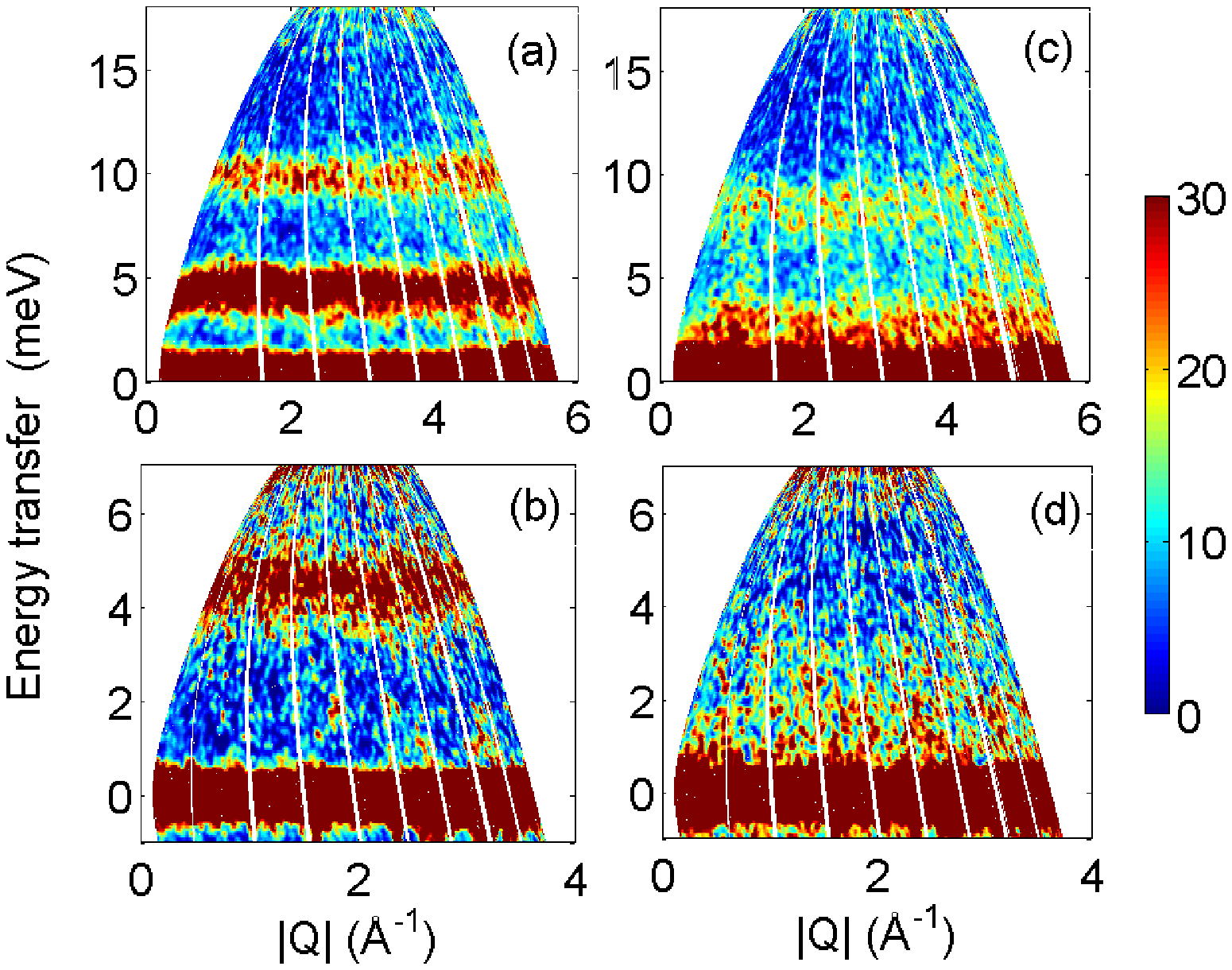}
\caption {(Color online) Contour plots of the inelastic scattering intensity (in mb/sr meV f.u.) plotted as energy transfer vs. wave vector transfer ($Q$) of CeNiC$_2$ measured with $E_i$ = 20 meV [(a) and (c)] and 8 meV [(b) and (d)] at 5 and 25 K respectively.}
\end{figure}

\begin{figure}[b]
\vskip 0.4 cm
\centering
\includegraphics[width = 4 cm]{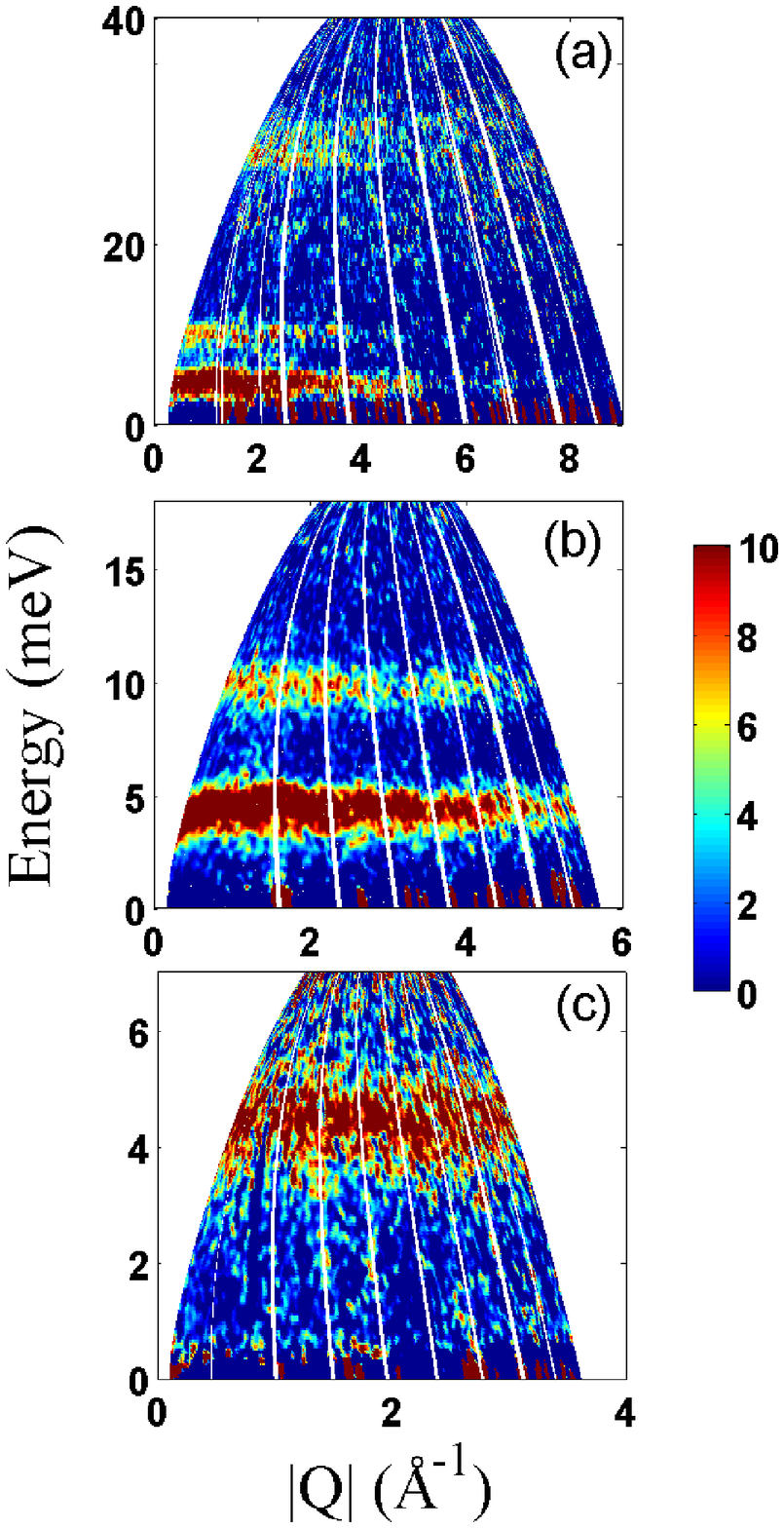}
\caption {(Color online) The color-coded contour map of the magnetic scattering of CeNiC$_2$ at 5 K estimated by subtracting (see text) the data of LaNiC$_2$ measured with incident energy $E_i$ = (a) 50 meV, (b) 20 meV  (c) and 8 meV respectively.}
\end{figure}

\section{RESULTS AND DISCUSSION}

\subsection{Room temperature Neutron diffraction}

Powder neutron diffraction measurements were carried out at room temperature on the polycrystalline RNiC$_2$ (R = Ce and La) samples. Rietveld refinement was carried out using the GSAS software on the basis of orthorhombic CeNiC$_2$ type crystal structure (Space group No. 38).~\cite{Kotsanidis} In this structure Ce/La atom occupies 2a sites, Ni atom 2b sites and C atom 4e sites.  Rietveld refinement profile along with neutron diffraction patterns of these compounds are shown in Figs. 1 (b)-(c). One small impurity peak was detectable in CeNiC$_2$ (¡« 1\% of the intensity of the maximum sample peak) whilst none were observed in LaNiC$_2$, indicating that the samples are  single phase. The site occupancies were all fixed at 100\%. The results of the refinements are displayed in Table I. Observed lattice parameters are in agreement with the previously reported values.~\cite{Kotsanidis} The nearest-neighbor distances are 3.8305 \AA ~ for Ce-Ce,  3.0671 \AA ~for Ce-Ni, 1.9687 \AA ~for Ni-C.

\subsection{Magnetization and Heat capacity}

The temperature ($T$) variation of the dc magnetic susceptibility ($\chi = M/H$, where $H$ is the applied magnetic field) measured in zero field cooled  condition in $H$ = 40 kOe is shown in Fig. 2 (a).  $\chi$ shows a drop  below 20 K with decreasing $T$. This corresponds to the PM/AFM transition ($T_{N_1}$) in the sample. The magnetic susceptibility of CeNiC$_2$ above 50 K exhibits Curie-Weiss behavior. A linear least-squares fit yields an effective magnetic moment $p_{eff}$= 2.30 $\mu_B$, which is close to free Ce$^{3+}$-ion value (2.54 $\mu_B$), and a negative paramagnetic Curie temperature $\theta_p$= $-$15 K. The value of magnetic moment suggests that the Ce atoms are in their normal Ce$^{3+}$ valence state, which agrees with smooth variation of the unit cell volume in RNiC$_2$ series.~\cite{Kotsanidis} Negative value of $\theta_p$ is indicative of a negative exchange constant and AFM ordering. LaNiC$_2$ shows  weak Pauli paramagnetism type behavior (not shown here, which confirms that Ni is nonmagnetic in RNiC$_2$ series).~\cite{Pecharsky}

\par
Fig. 2 (b) shows the $M$ versus $H$ isotherms recorded at different temperatures.  $M-H$ data imply that the net magnetization in the ordered state of CeNiC$_2$ is extremely low. It is far from saturation and barely reaches 7\% of that expected theoretically ~$gJ$= 2.14 $\mu_B$ for Ce$^{3+}$ ions, in a 70 kOe magnetic field. This is consistent with previously reported results.~\cite{K Motoya, W. Sch} The low values of the observed magnetization is expected for an AFM ground state due to the cancellation of magnetization from different magnetic sublattices of Ce ions.

\par

Zero field $C_P$ versus $T$ data of both compounds from 2 to 300 K are shown in Fig. 3 (a). For LaNiC$_2$, $C_P(T)$ data show sharp discontinuity between 2 and 3 K which is consistent with the transition to the superconducting state. The discontinuity in the heat capacity of LaNiC$_2$, $\Delta C/\gamma T_C$ = 1.20, suggests weak electron-electron coupling and confirms the bulk nature of superconductivity.~\cite{Pecharsky} For CeNiC$_2$ a clear signature of anomaly is observed at $T_{N_1}$ = 20 K and $T_{N_3}$ = 2.5 K and a further weak anomaly is observed at $T_{N_2}$ = 10 K which matches well with previous reports.~\cite{Pecharsky}  We have  carefully looked at low $T$ behavior of $C_P$.  At $T \ll \Theta_D$ ($\Theta_D$ = Debye temperature), the lattice part of the heat capacity $C_{debye}$  has a $T^3$ dependence. We estimate $\gamma$ and  $\Theta_D$ using same method as reported by V. K. Pecharsky et.al.~\cite{Pecharsky} and find  similar values of $\gamma$ (6.0 mJ mol$^{-1}$ K$^{-2}$) and  $\Theta_D$ (380 K).

\par
The magnetic contribution to the specific heat $C_{mag}$ is shown in Fig. 3 (b). The $C_{mag}$ was estimated by subtracting off the lattice contribution equal to the specific heat of isostructural LaNiC$_2$. The effect of crystal electric field is reflected as a broad Schottky-type anomaly centered around 52 K in $C_{mag}$. The solid curve in Fig. 3(b) represents the crystal field contribution to specific heat according to the CEF level scheme obtained from the analysis of inelastic neutron scattering data. The magnetic contribution to entropy $S_{mag}$ was obtained by integrating the $C_{mag}(T)/T$ versus $T$ plot and is shown in Fig. 3(c). A value of $S_{mag}$ around the magnetic ordering temperature is comparable with $Rln2$ suggesting a CEF split doublet ground state in CeNiC$_2$ and is confirmed by the INS data in section D.
\par

\subsection{Muon spin relaxation}
The time dependence of asymmetry ($\mu$SR) spectra of CeNiC$_2$ measured at various temperatures in zero field are shown in Figs. 4 (a)-(f). The spectra show exponential type decay above 20 K in the paramagnetic state. In the temperature range 10 K $\le T \le$ 20 K, muon spin precession with two frequencies is observed, indicating that at least two muon sites exist in the compound. Below 10 K, as shown in Fig. 4 (a), muon spin precession is not observable due to the fact that internal fields exceed the maximum internal field detectable on the $\mu$SR spectrometer due to the pulse width of the ISIS muon beam. In between these temperature ranges 2 K $\le T \le$ 10 and 20 K $\le T \le$ 30 K, we used simple exponential decay plus constant background to fit our $\mu$SR spectra. In the range 10 K $\le T \le$ 20 K, the spectra were fitted with,~\cite{Smidman2013,Hillier2012}

\begin{equation}
G_z(t) =\sum_{i=1}^{n} A_i cos(\gamma_\mu H_i t+\phi)e^{-\frac{{\sigma^2_i t}^2}{2}}+A_3 e^{-\lambda t} + A_{bg}
\end{equation}

\par
where the initial amplitude of the exponential decay and the oscillatory component are $A_3$ and $A_i$ respectively, the internal magnetic fields at the muon stopping site $i$ are $H_i$, the Gaussian decay rate is $\sigma_i$, $\lambda$ is the muon depolarization rate, $\phi$ is the common phase, $\gamma_\mu/2\pi$ = 135.53 MHz T$^{-1}$ and $A_{bg}$ is the background. The temperature dependencies of these parameters are shown in Figs. 5 (a)-(d). At 20 K, as shown in Fig. 5 (a) there is a loss of 2/3 value of the initial asymmetry (1/3 is left) of $A_3$ from the high temperature value. The initial asymmetry associated with frequency terms $A_1$ and $A_2$ start to increase below this temperature (20 K) [see Fig. 5 (a)], indicating the presence of a long-range ordered state in CeNiC$_2$ which agrees with the specific heat, magnetic susceptibility and neutron diffraction data.~\cite{Onodera1998, K Motoya, W. Sch, M. Murase, Pecharsky} The temperature dependence of the exponential decay term is shown in Fig. 5 (b). The muon depolarization rate ($\lambda$) was found to suddenly increase at $T_{N_1}$, indicating a transition between the paramagnetic and ordered states. However $\lambda$ shows a weak anomaly at $T_{N_2}$ and $T_{N_3}$ where there is a rearrangement of the spins and a change in the magnetic structure which is in agreement with the reported change in the propagation vector from the neutron diffraction study.~\cite{motoya} Fig. 5(c) shows the temperature dependence of the muon depolarization rate. $\sigma_1$ and $\sigma_2$ remain almost constant within the temperature range 10 K $\le T \le$ 20 K.

In order to find out the nature of the magnetic interaction in CeNiC$_2$ one of the two temperature dependence of internal fields was fitted with

\begin{equation}
H_1(T)= H_0\left(1-\left(\frac{T}{T_N}\right)^{\alpha}\right)^{\beta}
\end{equation}

\par

Observed parameters are $\beta$= 0.31, $H_0$ = 478 G, $\alpha$ = 3.12 and $T_{N}$ = 19.1 K (see Fig. 5 (d)). A good fit with $\beta$ = 0.31 suggests the magnetic interactions in CeNiC$_2$ are 3D Ising spin system with long-range spin-spin interactions. $\alpha >$  1 indicates complex magnetic interactions in this system.

\begin{figure}[b]
\vskip 0.4 cm
\centering
\includegraphics[width = 7.5 cm]{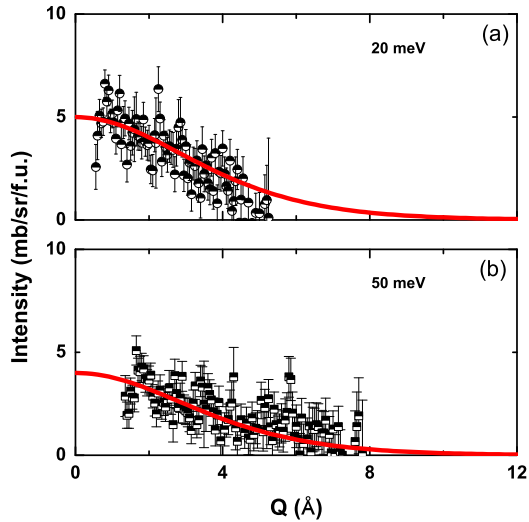}
\caption {(Color online) The Q dependence of total energy integrated intensity  between (a) 5.85 and 11.05 meV and (b)  22 and 35 meV at 25 K for incident energy $E_i$ =20 meV and 50 meV respectively. The solid line represents the square of the Ce$^{3+}$ magnetic form factor (scaled to matching with the data).}
\end{figure}

\begin{figure}[t]
\vskip 0.4 cm
\centering
\includegraphics[width = 9.3 cm]{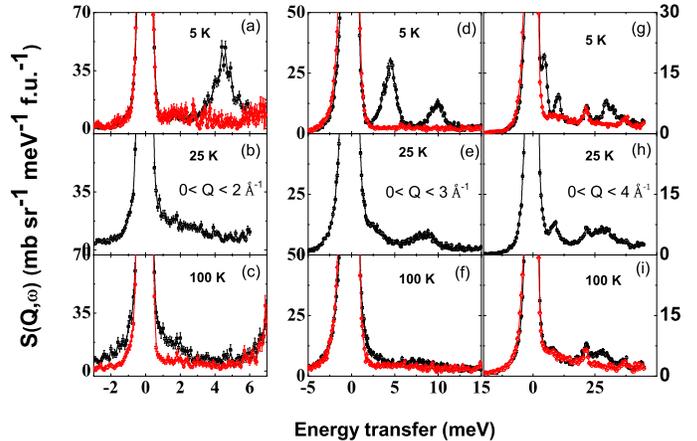}
\caption {(Color online) The Q-integrated 1D cuts of the total scattering from CeNiC$_2$ (black square) and LaNiC$_2$ (red circle) with an incident energy $E_i$ = 8 meV (left panel), 20 meV (middle panel) and 50 meV (right panel) respectively.}
\end{figure}

\par
We also recorded the field dependent asymmetry spectra at 1.2 K and fitted them using an exponential function plus constant background [see Fig. 6 (a)]. Initial asymmetry ($A_3$) and depolarization term ($\lambda$) as a function of applied longitudinal field at 1.2 K is shown in Figs. 6 (b)-(c). The relaxation rate decreases with applied field as is conventionally expected, and we fitted the field dependence using a modified version of Redfield’s equation,~\cite{smd} 

\begin{equation}
\lambda = \lambda_0+\frac{2\gamma^2_{\mu}\Delta^2\tau}{1+\gamma^2_{\mu}H^2\tau}
\end{equation}

where $\gamma_{\mu}$ is the muon gyromagnetic ratio, $\Delta$ describes the width of the field distribution, and $\tau$ is the characteristic timescale for the spin fluctuations experienced by the muons. The fit of $\lambda(H)$ data by Eq. (3) is shown by solid red curve in Fig. 6 (c). A good fit is obtained for $\lambda_0$= 0.002 $\mu s^{-1}$, $\Delta$= 3.6 G, and $\tau_C$= 20$\times$ 10$^{-8}$ s. The observed value of $\tau_C$ in CeNiC$_2$ is comparable with CePd$_{0.15}$Rh$_{0.85}$.~\cite{adroja} 

\par
Field dependence of $A_3(H)$ (with $\lambda$ as a free parameter) is shown in Fig. 6 (b). We used a quadratic-type decoupling function which is appropriate for a Lorentzian distribution of fields,~\cite{smd} 

\begin{equation}
A_3(H)=f_0+F \frac{b^2}{1+b^2}
\end{equation}

where the meaning of $f_0$, $F$ and $b$ (= $H/H_0$, $H_0$ internal field) parameters are given in Ref. 21.~\cite{smd} A fit has been made using Eq. (4) and an internal field of 1250(40) G was obtained. The larger value of the internal field observed at 1.2 K compared to that estimated between 10 and 20 K again indicates different magnetic structures at 1.2 K and in between 10$-$20 K.

\subsection{Inelastic Neutron Scattering}

In order to understand the origin of the reduced moment magnetism and also to investigate crystal field excitations and their energy level scheme in CeNiC$_2$ we have performed an inelastic neutron scattering study. Neutrons with incident energies $E_i$ = 8, 20, and 50 meV were used to record the INS spectra at 5, 25, 50 and 100 K for scattering angles between 3$^{\circ}$ and 135$^{\circ}$. The INS measurements were carried out on the polycrystalline samples of CeNiC$_2$ and LaNiC$_2$ using the MARI spectrometer. The data of LaNiC$_2$ were used to subtract the phonon contribution in CeNiC$_2$.

\par

For magnetic neutron scattering the partial differential cross section, which measures the probability of scattering per solid angle per unit energy, is~\cite{lovesey}

\begin{equation}
\frac{d^2\sigma}{d\Omega dE'}= \frac{k'}{k}\frac{N}{\hbar}(\gamma r_0)^2F(Q)^2\sum_{\alpha \beta}(\delta_{\alpha\beta}-\widehat Q_{\alpha}\widehat Q_{\beta})S^{\alpha\beta(Q,\omega)}
\end{equation}

where $k^{'}$ and $k$ are the scattered and incident  neutron wavevectors, $\gamma r_0$ = 5.391 fm is the magnetic scattering length, $Q$ is the momentum transfer, $F(Q)$ is the magnetic form factor,  $\omega$ is the energy transfer,  $N$ is the number of moments,  and the summation runs over the Cartesian directions. The magnetic scattering function $S^{\alpha\beta(Q,\omega)}$ is proportional to the space and time Fourier transform of the spin-spin correlation function.~\cite{lovesey}
\par

To determine the correct magnetic contribution to the measured spectra, we have subtracted the phonon contribution using the data of nonmagnetic reference compound. We first used a direct subtraction method [$S(Q,\omega)_{CeNiC_2}-S(Q,\omega)_{LaNiC_2}$] and then after allowing for a difference in the total scattering cross section [$S(Q,\omega)_{CeNiC_2}-\sigma_c\times S(Q,\omega)_{LaNiC_2}$] (where $\sigma_c$= 0.8288 is the ratio of the total scattering cross-section of CeNiC$_2$ and LaNiC$_2$).~\cite{Adroja2011, Adroja2012} Both analyses gave similar magnetic response. But the second method gave slightly better estimation at low energy side in subtracting the nuclear elastic peak at zero energy transfer and hence in this paper we present all our data analyzed using the second method. 

\par

Figs. 7 (a)-(d) display the color-coded plot of the scattering intensity of energy vs.\ momentum transfer of CeNiC$_2$. Two inelastic excitations at 4.5 and 10 meV were observed with a significant intensity at low scattering vectors at 5 and 25 K. Absence of these excitations in LaNiC$_2$ indicates they are magnetic in origin. 5 K data has an additional magnetic excitation with a maximum at around 4.5 meV. This excitation is not present at 25 K as shown in Figs. 7 (b)-(d) where the elastic line is broadened signifying the presence of spin wave below magnetic ordering. In the paramagnetic state, the spectral weight is shifted towards the elastic line, and quasielastic scattering (QES) is observed. Figs. 8 (a)-(c) reveals the color-coded contour map of the magnetic scattering of CeNiC$_2$ at 5 K estimated by subtracting the data of LaNiC$_2$ (using second method) measured with incident energies $E_i$= 50 meV (a), 20 meV (b) and 8 meV (c) respectively.

\par

Figs. 9 (a) and (b) represent the $Q$-dependent energy integrated intensity between (i) 5.85 and 11.05 meV , (ii) 22 and 35 meV at 25 K of CeNiC$_2$ for incident energy $E_i$ =20 and 50 meV respectively. It follows the square of Ce$^{3+}$ magnetic form factor [$F^2 (Q)$], which suggests that the inelastic excitations result mainly from single-ion CEF transitions. The scattering at the highest $Q$ (not shown here) is comparable for CeNiC$_2$ and LaNiC$_2$, which indicates similar phonon contributions in these compounds. At low $Q$, the magnetic scattering is strong and with a small phonon contribution in CeNiC$_2$. This can be seen clearly in the 1D cuts made from the 2D color plots at low $Q$ from 0 to 4 \AA$^{-1}$ (see Fig. 10) at 5, 25 and 100 K. 

\par

\begin{figure}[h!]
\vskip 0.4 cm
\centering
\includegraphics[width = 8 cm]{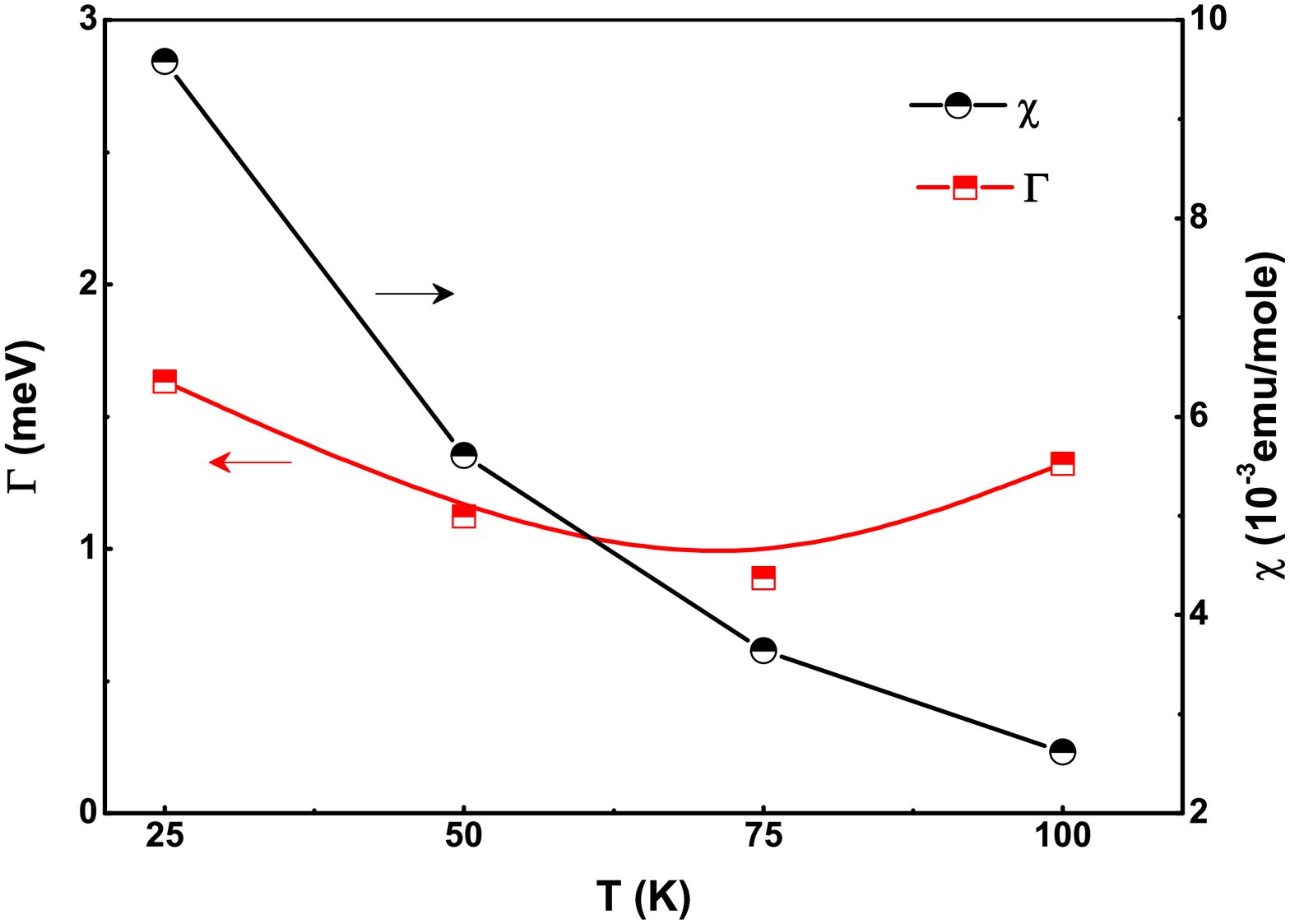}
\caption {(Color online) Temperature dependence of the quasielastic linewidth (left y-axis) and susceptibility (right y-axis) obtained from fitting INS data measured with an incident energy of 8 meV.}
\end{figure}

\par

\begin{figure}[t]
\vskip 0.4 cm
\centering
\includegraphics[width = 9 cm]{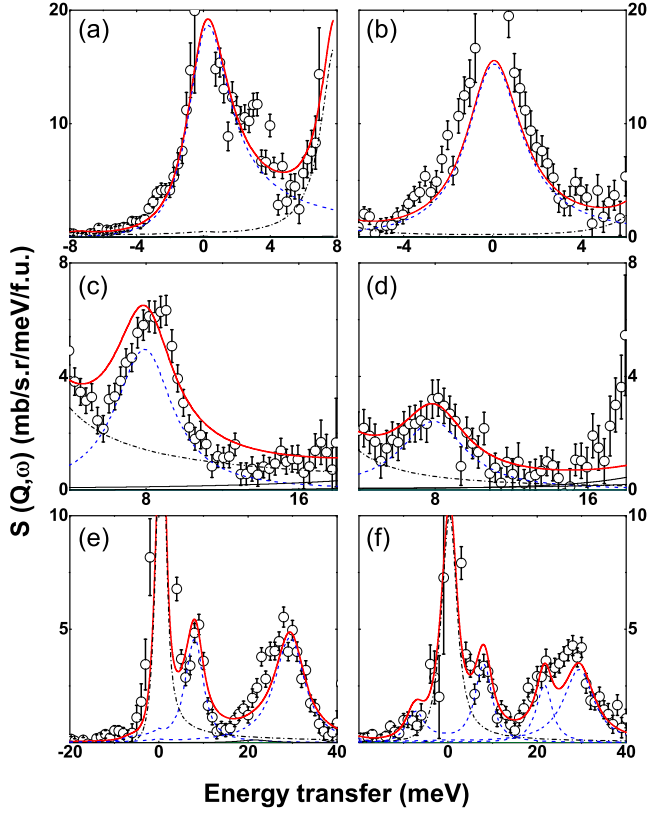}
\caption {(Color online) The estimated magnetic scattering of CeNiC$_2$ at 25 K (left panel) and 100 K (right panel) for momentum transfer $|Q|$ from 0 to 2 \AA$^{-1}$ for 8 meV, $|Q|$ from 0 to 3 \AA$^{-1}$ for 20 meV and $|Q|$ from 0 to 4 \AA$^{-1}$ for 50 meV. The solid lines are the fits based on the crystal electric field model and the dashed and dash-dotted lines are the components of the fit.}
\end{figure}

\par

In order to estimate the quasielastic linewidth we have analyzed the 8 meV data using a Lorentzian function. We used an elastic line resolution function and an additional Lorentzian function to model the quasielastic component to fit our data. The widths of the elastic component were fixed from the measurements of vanadium with the same incident energy and frequency of the Fermi chopper. Fig. 11 (left y-axis) shows the $T$ dependence of the half width at half maximum (HWHM, $\Gamma$). $\Gamma$ doesn't change much with temperature.  An estimate of the Kondo temperature ($T_K$) can be obtained from the value of $\Gamma$ at 0 K (we used $\Gamma$ value in between 50 and 80 K). For CeNiC$_2$, $\Gamma$ value gives an estimation of  $T_K$ = 11 K. However at lower temperature (around 25 K) a slightly higher $\Gamma$ value was observed. We attribute this to the presence of short range magnetic correlations just above $T_{N_1}$.

\par

We now proceed with a more detailed analysis of the observed CEF excitations. In the orthorhombic point symmetry, Amm2 ($C_{2v}$), at the Ce$^{3+}$ site and taking the $z$-axis as the quantization axis (i.e. $x\|a$, $y\| b$ and $z\|c$), the CEF Hamiltonian can be represented as follows:

\begin{equation}
H_{CEF} = B^{0}_{2}O^{0}_{2}+B^{2}_{2}O^{2}_{2}+B^{0}_{4}O^{0}_{4}+B^{2}_{4}O^{2}_{4}+
B^{4}_{4}O^{4}_{4}
\end{equation}

where $O^{m}_{n}$ are the Stevens operators and  $ B^{m}_{n}$ are the CEF parameters to be determined from the experimental data of inelastic neutron scattering.~\cite{Stevens1952, Stevens1967} The value of $ B^{0}_{2}$ and $ B^{2}_{2}$  can be accurately determined using the high-temperature expansion of the magnetic susceptibility,~\cite{jensen, Yutaka} which gives $ B^{0}_{2}$ and $ B^{2}_{2}$ in terms of the CW temperatures, $\theta_a$ and $\theta_b$ for an applied field parallel to $a$-axis and $b$-axis, respectively and $B^{0}_{2}$ in terms of $\theta_c$ for an applied field parallel to $c$-axis. The detailed formulation have been given in ref [9]. From their formula we have determined the value of $B^{0}_{2}$ = 0.08531  meV and $B^{2}_{2}$ = -0.8913 meV. It is to be noted that this values are valid for an isotropic exchange interaction. They were kept fixed in our initial analysis of the INS data which did not give good fit to the data and then finally we allowed to vary all five CEF parameters independently.

\begin{table}
\begin{center}
\caption{CEF parameters $B^{m}_{n}$, molecular field parameters ($\lambda_\xi$ , $\xi= a, b, c$ ) and temperature-independent constant susceptibility ($\chi_{\xi}$). The parameters were estimated by simultaneous fit to the INS data at 25 K and 100K and then fitting  the single-crystal susceptibility.}
\begin{tabular}{lcccccccccccc}
\hline
    &&   && 25  and 100 K  \\   
\hline
\hline
$B^{0}_{2}$ (meV)  &&  && 0.080(2)  &&       		 \\ 					
$B^{2}_{2}$ (meV) && &&-0.650(1) &&         		\\ 
$B^{0}_{4}$ (meV) && &&-0.086(2)&&        		 \\ 
$B^{2}_{4}$ (meV)  && &&-0.050(3) &&          	\\ 
$B^{4}_{4}$ (meV)  &&   &&0.209(2) &&        		 \\ 
$\lambda_a$ (mole/emu)  && &&3.856 &&          \\ 
$\lambda_b$ (mole/emu)  && && -51.214 &&      \\ 
$\lambda_c$ (mole/emu)  && && -7.234 &&         \\ 
$\chi_a$ ($\times 10^{-3}$ emu/mole)  && && 2.505&&       \\ 
$\chi_b$ ($\times 10^{-3}$ emu/mole)  && && 2.626 &&        \\ 
$\chi_c$ ($\times 10^{-3}$ emu/mole) &&  && 2.642&&       \\ 
\hline 
\hline
\end{tabular}
\end{center}
\end{table}

\par
The solid lines in Figs. 12 (a)-(f) represent the fit to the CEF model for simultaneous refinement of all six data sets for 8, 20, and 50 meV incident energies (we first refined 25  and 100 K data). The phenomenological crystal field parameters $B^{m}_{n}$ obtained from the best fit are given in table II.

The CEF wave functions obtained from the simultaneous fit to 25 and 100 K, INS data are given by,

\begin{equation}
|\psi^{\pm}_{1}>=0.4104|\pm \frac{5}{2}>+0.9115|\pm \frac{1}{2}>-0.0273| \pm \frac{3}{2}> \\
\end{equation}
\begin{equation}
|\psi^{\pm}_{2}>=0.8719|\pm \frac{5}{2}>-0.4010|\pm \frac{1}{2}>-0.2812| \pm \frac{3}{2}>\\
\end{equation}
\begin{equation}
|\psi^{\pm}_{3}>=0.2672|\pm \frac{5}{2}>-0.0916|\pm \frac{1}{2}>+0.9593| \pm \frac{3}{2}>\\
\end{equation}

$\psi_1$, is the GS wave function, whereas,  $\psi_2$ is for the first excited state at 92 K and  $\psi_3$ is for the second excited state at 342 K above the GS. The GS magnetic moments of the Ce atoms  $<\mu_x>$, $<\mu_y>$ and $<\mu_z>$, along the three crystallographic axes $a$, $b$ and $c$, respectively can be calculated from

\begin{equation}
<\mu_z>=<\psi^{\pm}_{1}|g_JJ_z|\psi^{\pm}_{1}> \\
\end{equation}
\begin{equation}
<\mu_x>=<\psi^{\pm}_{1}|\frac{g_J}{2}(J^++J^-)|\psi^{\pm}_{1}> \\
\end{equation}
\begin{equation}
<\mu_y>=<\psi^{\pm}_{1}|\frac{g_J}{2i}(J^+-J^-)|\psi^{\pm}_{1}> \\
\end{equation}

Calculated magnetic moments in presence of an applied magnetic field ($H$ = 10 kOe; $T$ = 1 K) using GS wave function are $<\mu_x>$ = 0.583 $\mu_B$, $<\mu_y>$ = 0.6876 $\mu_B$ and $<\mu_z>$ = 0.341 $\mu_B$. This is in agreement with neutron diffraction data, which gives possible direction of moment along $b-$axis.~\cite{motoya} However it is to be noted that the neutron diffraction study was unable to find accurate direction and magnitude of the moment due to limited numbers of magnetic Bragg reflections.~\cite{motoya}

\par
\begin{figure}[t]
\vskip 0.4 cm
\centering
\includegraphics[width = 7 cm]{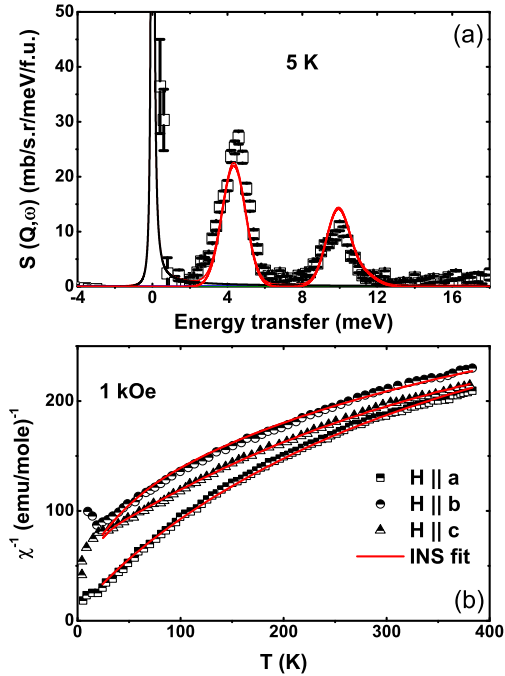}
\caption {(Color online) (a) The estimated magnetic scattering of CeNiC$_2$ at 5 K with $E_i$=20 meV. The solid line represents the fit using CEF and molecular field (see text). (b) Temperature dependence of the inverse susceptibilities along the three crystallographic axes of CeNiC$_2$. The solid line represents the fit based on the CEF model.}
\end{figure}

\par

After obtaining CEF parameters from the simultaneous fit of  the six INS data sets at 25 and 100 K, we have analyzed 20 meV INS data at 5 K (i.e. below the magnetic ordering) adding the molecular field term in the Hamiltonian. We kept the values of CEF parameters  (which were obtained from 25 and 100 K as mentioned above) fixed for the analysis of 5 K data and varied molecular fields. First we varied only molecular field along $x-$axis ($H_x$), while keeping zero-molecular fields along $b-$ ($H_y$) and $c-$axes ($H_z$). This method did not give good fit to the data. Then we tried to vary $H_y$ (while keeping $H_x$=0 and $H_z$=0), however we did not obtain good fit for this method either. Hence finally we varied $H_x$ and $H_y$ simultaneously (keeping $H_z$=0), which gave a reasonably good fit to the data (it is to be noted that when we also allowed to vary $H_z$ it gave almost zero value). The values of the molecular field estimated from the analysis are $H_x$= 79.3(1.8) T and $H_y$=76.5(1.7) T. The quality of the fit can be seen in Fig. 13 (a). The presence of molecular fields along $a-$ and $b-$axes may suggests that the Ce moment is confined to the $ab$-plane and not along $a-$ or $b-$ axis. A detail neutron diffraction investigation using a high neutron flux instrument on CeNiC$_2$ single crystal will be helpfully to understand the presence of the molecular field along $a$ and $b$-axes. To check reliability of our set of CEF parameters we model the single crystal magnetic susceptibility data from literature.~\cite{Onodera1998} Observed anisotropic magnetic susceptibility data correctly fits with our calculated $\chi(T)$ using CEF parameters [see Fig. 13 (b)].

\section{Conclusions}

We have studied the magnetic properties of CeNiC$_2$ using magnetic susceptibility, heat capacity, inelastic neutron scattering and  $\mu$SR measurements. A magnetic transition around 20 K is observed with the emergence of oscillations in zero field $\mu$SR spectra. We fitted the temperature dependence of the internal magnetic fields to a model of a mean field magnet, which revealed 3D nature of magnetic interactions. 

\par

INS measurements of polycrystalline CeNiC$_2$  at low temperatures indicate two CEF excitations at 8 and 30 meV. At 5 K, we observe an additional peak at 4.5 meV due to spin-wave excitations. Above $T_N$, this peak is not present, but quasielastic scattering is observed. A linear fit to the temperature dependence of the quasielastic linewidth gives an estimate of $T_K$ = 11 K. From an analysis of INS and magnetic susceptibility data with a CEF model, we propose a CEF scheme for CeNiC$_2$. The CEF scheme correctly predicts the direction of the ordered moment, but the observed magnetic moment at 2 K  is 0.6876 $\mu_B$/Ce which is higher compared to the moment as observed from neutron diffraction (0.25 $\mu_B$/Ce). We believe that the observed reduced moment is due to presence of  hybridization between the localized Ce$^{3+}$ $f$-electrons and the conduction band. Further the susceptibility analysis reveals the anisotropic molecular fields, stronger along the $b$-axis, which is also in agreement with the direction of moment along the $b$-axis. The spin wave measurements on CeNiC$_2$ single crystal will help to shade more light on the anisotropic exchange interactions in CeNiC$_2$.

\par
We note furthermore that the estimated on-site Kondo exchange (11 K) is of the same order of magnitude as the magnetic phase transition temperatures, which classifies this compound as a magnetically ordered Kondo lattice. The complexity of spin ordering in CeNiC$_2$ is inferred as a consequence of admixing of crystal electric field energies together with Kondo and RKKY exchange, and we believe that our study and determination of CEF parameters presented in this work establishes the essential ingredients with which to formulate an understanding of the ground state in CeNiC$_2$.

\section{Acknowledgment}
AB thanks the FRC of UJ and ISIS-STFC for funding support. DTA and ADH would like to thank CMPC-STFC, grant number CMPC-09108, for financial support.  AMS thanks the SA-NRF (Grant 78832) and UJ Research Committee for financial support.

\end{document}